# Generation and Hall effect of skyrmions enabled via using nonmagnetic point contacts

Zidong Wang[1,*], Xichao Zhang[2,*], Jing Xia[2,*], Le Zhao[1], Keyu Wu[1], Guoqiang Yu[3], Kang L. Wang[4], Xiaoxi Liu[5], Suzanne G. E. te Velthuis[6], and Axel Hoffmann[6], Yan Zhou,[2] and Wanjun Jiang[1,†]

[1]*State Key Laboratory of Low-Dimensional Quantum Physics and Department of Physics, Tsinghua University, Beijing 100084, China*
[2]*School of Science and Engineering, The Chinese University of Hong Kong, Shenzhen, Guangdong 518172, China*
[3]*Beijing National Laboratory for Condensed Matter Physics, Institute of Physics, Chinese Academy of Sciences, Beijing 100190, China*
[4]*Department of Electrical Engineering, University of California, Los Angeles, California, 90095, USA*
[5]*Department of Electrical and Computer Engineering, Shinshu University, 4-17-1 Wakasato, Nagano 380-8553, Japan*
[6]*Materials Science Division, Argonne National Laboratory, Lemont, Illinois, 60439, USA*

*Author Contributions

Zidong Wang, Xichao Zhang and Jing Xia contributed equally to this work.

†Corresponding Author

E-mail: jiang_lab@tsinghua.edu.cn

**To enable functional skyrmion-based spintronic devices, the controllable generation and manipulation of skyrmions is essential. While the generation of skyrmions by using a magnetic geometrical constriction has already been demonstrated, this approach is difficult to combine with a subsequent controlled manipulation of skyrmions. The high efficiency of skyrmion generation from magnetic constrictions limits the useful current density, resulting in stochastic skyrmion motion, which may obscure topological phenomena such as the skyrmion Hall effect. In order to address this issue, we designed a nonmagnetic conducting Ti/Au point contact in devices made of Ta/CoFeB/TaO$_x$ trilayer films. By applying high voltage pulses, we experimentally demonstrated that skyrmions can be dynamically generated. Moreover, the accompanied spin topology dependent skyrmion dynamics – the skyrmion Hall effect is also experimentally observed in the same devices. The creation process has been numerically reproduced through micromagnetic simulations in which the important role of skyrmion-antiskyrmion pair generation is identified. The motion and Hall effect of the skyrmions, immediately after their creation is described using a modified Thiele equation after taking into account the contribution from spatially inhomogeneous spin-orbit torques and the Magnus force. The simultaneous generation and manipulation of skyrmions using a nonmagnetic point contact could provide a useful pathway for designing novel skyrmion based devices.**

## Introduction

Magnetic skyrmions are topological spin textures that can be found in chiral bulk magnets[1-4] and asymmetric multilayers[5-16]. In terms of potential spintronic applications, magnetic multilayers are particularly interesting not only for providing ample material systems hosting room-temperature skyrmions, but also for enabling efficient electrical manipulations through current-induced spin-orbit torques[10,14,15,17-20]. These asymmetric multilayers are typically made of inversion symmetry breaking trilayers with a perpendicular magnetic anisotropy, which contain an interfacial, noncollinear Dzyaloshinskii-Moriya interaction (DMI)[21-24] that stabilizes Néel-type skyrmions with a fixed spin chirality, in contrast to swirling Bloch-type skyrmions in most bulk chiral magnets[25]. As such, many interesting topological physics and device concepts have been revealed with direct relevance for spintronic storage and logic applications[18,26-29]. On the other hand, as skyrmions may become key components for future data storage and logic devices, electrical generation and manipulation in a controllable manner are thus essential[30,31].

The generation of magnetic skyrmions has recently been demonstrated by applying magnetic fields[10], spin-polarized currents[27,32], local heating[33], laser beams[34,35] and electric voltage[36,37]. Meanwhile, skyrmions can also be generated in geometrical constricted devices[6,38-42] by applying electrical currents[43]. For example, the dynamical generation of skyrmions can be realized by squeezing chiral band domains through a (magnetic) geometrically constriction after being subjected to spatially inhomogeneous spin-orbit torques (SOTs)[6,40,41,44,45]. In addition, skyrmions can also be generated in a constricted region as a result of inhomogeneous current induced nonlinear magnetization dynamics[38,41]. Similarly, magnetic inhomogeneities with a homogeneous current may result in skyrmion nucleation[46]. However, the simultaneous occurrence of skyrmion generation and spin-topology dependent motion − skyrmion Hall effect[11,16,42,47-49] in devices with geometrical constrictions, has not been experimentally demonstrated. The skyrmion Hall effect is important for identifying the topological nature of Néel-type skyrmions found in the related magnetic multilayers, and also provides a basis for enabling electrically programmable skyrmionic devices. Early micromagnetic simulations suggested two possible regimes for skyrmion generation[44]. At low current, skyrmion nucleation may proceed via deformation of existing domain walls in a process analogous to a Rayleigh-Taylor instability[6], while larger currents may provide large enough SOTs to eliminate pre-existing magnetic domain structures and nucleate new skyrmions from inhomogeneous SOTs. Since this latter mechanism does not require pre-existing magnetic domains, we explore in this work the use of inhomogeneous current injection from a nonmagnetic point contact.

So far, it is known that the skyrmion Hall effect occurs when the translational motion of skyrmion is in the steady flow motion region, which demands a sufficiently large driving current density[6,31,50]. While typical geometrically constricted devices facilitate the formation of skyrmions, they can only tolerate a relatively small current density, above which the constriction will break[6,41]. In the smaller current density region, the translational motion of skyrmion is strongly influenced by the randomly distributed pinning sites inside the materials. Namely, skyrmions stochastically hop from one pinning site to another resulting in creep motion. This explains the previous inconclusive results with respect to the skyrmion Hall effect in the devices with magnetic constrictions[6,41]. In the present work, we will demonstrate the simultaneous observation of skyrmion generation and the skyrmion Hall effect in modified Ta(5nm)/CoFeB(1.1nm)/TaO$_x$(3nm) devices with a nonmagnetic point contact. Our experimental observation is made possible by replacing the previous thin resistive magnetic geometrical constriction Ta/CoFeB/TaO$_x$ with a conducting nonmagnetic Ti/Au point contact.

The choice of a thick Ti (10 nm)/Au(100 nm) electrode allows sufficiently large electric currents that can generate skyrmions through the divergence of current-induced SOTs, with less chance of the device failure and current densities sufficiently large for sustaining flow motion of the newly generated skyrmions. More importantly, it can also produce a steady flow motion of skyrmions together with less pronounced joule heating. It should be mentioned here that the usage of a nonmagnetic point contact excludes the generation of skyrmions inside the contact[6,38,39,41]. The location of the skyrmion generation should occur at the contact region between the constriction and FM trilayer where the divergence of the current induced SOTs is maximized.

We start to formulate the working mechanism of the proposed device includes a narrow conducting nonmagnetic Ti/Au point contact, as shown in Fig. 1a. After injecting an electron current $+j_e$ (from left to right), the constricted device naturally introduces a spatially inhomogeneous current distribution, as shown in Fig. 1b. The convergent (left)/divergent (right) distribution of electron current density $j_e$ in our device can be computed by solving the Laplace's equations with boundary conditions of fixed potentials at the left and right ends of the device[44]. It can also be seen from Fig. 1b that $j_e$ is relatively large near the contact region. It should be noted that the electrostatic potential is assumed to be constant through the thickness of the device as the thickness of the FM/HM bilayer is very thin. In addition, the FM and HM layers exhibit comparable resistivities, and the thickness of the HM layer ($d_{HM}$) is five times larger than that of the FM layer ($d_{FM}$). Thus, the electron current is mostly flowing in the HM layer. Note that the direction of electron current $j_e$ is opposite to that of charge current $j_c$.

Heavy metals exhibit a strong spin-orbit interaction that results in a spin dependent preferential scattering, which is also known as the spin Hall effect[51,52]. Namely, in the FM/HM bilayer, a charge current flowing in the HM layer produces a vertical (transverse) spin current $j_s$, which gives rise to SOTs on the adjacent FM layer that can be used for manipulating the magnetization dynamics[53]. For our constricted device, the spin current distribution $j_s$ and its polarization for positive electron current flow is shown in Fig. 1c, which follows the same non-uniform distribution as charge current $j_e$ in the HM layer. It is worth mentioning that a small portion of the electron current directly flowing in the FM layer could theoretically induce magnetization dynamics through a spin-transfer torque. The contribution from spin-transfer torque to the overall skyrmion dynamics[27,44] has been estimated to be negligible, and therefore excluded in the present simulation.

**Micromagnetic simulation of skyrmion generation by using a nonmagnetic point contact.**

In order to reveal the feasibility of generating skyrmions using a nonmagnetic Ti/Au point contact, we first performed micromagnetic simulations by using the Object Oriented MicroMagnetic Framework (OOMMF)[54]. Following the modified Landau-Lifshitz-Gilbert (LLG) equation, the magnetization dynamics driven by SOTs can be written as:

$$\frac{d\bm{m}}{dt} = -\gamma_0 \bm{m} \times \bm{h}_{\text{eff}} + \alpha \left( \bm{m} \times \frac{d\bm{m}}{dt} \right) - \tau_{ad} [\bm{m} \times (\bm{m} \times \bm{p})] \tag{1}$$

where $t$ is time and $\alpha$ is the Gilbert damping coefficient. $\bm{h}_{\text{eff}} = -\delta\varepsilon/(\mu_0 M_S)\delta\bm{m}$ is the effective field that correlates with the functional derivative of the micromagnetic energy density $\varepsilon$ which includes the exchange, anisotropy, Zeeman, demagnetization and DMI terms. $\bm{m} = \bm{M}/M_S$ stands for the normalized vector of the magnetization with saturation magnetization $M_S$, $\bm{p}$ is the local spin-polarization direction, which is along $\hat{\bm{j}}_e \times \bm{n}$ in our device, where $\hat{\bm{j}}_e$ is the local electron current vector and $\bm{n}$ is the normal vector of the FM/HM interface. $\tau_{ad} = \left| \frac{\gamma_0 \hbar}{2\mu_0 e} \right| \frac{j_e \theta_{sh}}{d_{FM} M_S}$ represents the antidamping-like spin torque coefficient, where $\gamma_0$ is the gyromagnetic ratio, $\hbar$ is the reduced Planck constant, $\mu_0$ is the vacuum permeability, $e$ is the electron charge and $d_{FM}$ is the thickness of ferromagnetic layer. The spin Hall angle $\theta_{sh} = j_s/j_e$ measures the conversion efficiency from charge current to spin current. Through micromagnetic simulation, it is found that skyrmions can be generated even in the absence of field-like torque, we therefore only consider in the following the dominant contribution from the antidamping-like torque.

The device scheme for micromagnetic simulation mimics a device in which the HM layer, the FM layer, and nonmagnetic constricted electrode are built in a bottom-up fashion, as illustrated in Fig. 1(a). The FM and HM layers underneath the nonmagnetic electrode are completely etched away. The length and width of the FM and HM layers are $l = 3000$ nm and $w = 700$ nm, respectively. The nonmagnetic electrode with a length of $l_E = 200$ nm and a width of $w_E = 20$ nm is placed at the top center of the device, bridging the left and right halves of the FM layer. The thicknesses of the HM layer, FM layer, and electrode are equal to $d_{\text{HM}} = 5$ nm, $d_{\text{FM}} = 1.1$ nm, and $d_E = 5$ nm, respectively. The material specific parameters employed in our simulation are available in the Method. The spin configuration along the edge of the FM layer are fixed to the initial relaxed direction to prevent the unwanted destruction of skyrmions on the edge during the simulation. In order to reveal the topological properties of the generated magnetization textures, we have also calculated the evolution of the topological charge based on the magnetization configuration simulated by OOMMF. The topological charge of spin textures, namely, the skyrmion number, is expressed as $Q = 1/4\pi \int \boldsymbol{m} \cdot \left(\partial \boldsymbol{m}/\partial x \times \partial \boldsymbol{m}/\partial y\right) dxdy$. A well-defined skyrmion has a skyrmion number of value $Q = \pm 1$.

The numerical demonstration of the generation process of a single isolated skyrmion is enabled by using the spatially divergent antidamping-like SOT near the contact region. The electron current pulse is $j_e = 6$ MA cm$^{-2}$ that is normalized by the width of the wider part of device $w$ = 700 nm. Figure 2(a) illustrates the complete generation process of an isolated $Q = -1$ skyrmion, where the upper and lower panels show, respectively, the detailed evolutions of magnetization configurations and the associated topological charge densities [$i.e.$, $q(t) = \boldsymbol{m} \cdot \left(\partial \boldsymbol{m}/\partial x \times \partial \boldsymbol{m}/\partial y\right)$ ] as a function of time ($t$). It is clear that in the spin topology of the magnetization configuration there exists a transition from a topological trivial bubble with $Q = 0$ to a topological nontrivial skyrmion with $Q = -1$ [Fig. 2(b)]. In the first $t = 0\sim0.5$ ns (immediately after applying a current), the magnetization orientation near the contact region between constriction and trilayer reverses from upward to downward and forms a circular bubble domain with a substantial in-plane magnetization component aligned along the $+x$ direction. The associated topological charge can be calculated as $Q = 0$. This topologically trivial magnetic bubble continues to expand and moves into wider area of the FM layer driven by the spatially divergent current-induced SOTs, while $t = 0.5\sim1.0$ ns. During $t = 1.0\sim1.5$ ns, the in-plane magnetization component of the circular bubble domain rotates in-plane evolving into an outward configuration, forming an intriguing defect-like magnetization configuration. Now, its associated topological charge $Q = +1$ is calculated based on the magnetization profile, which is consistent with that of an antiskyrmion [indicated by the dashed circle in Fig. 2(a)].

In order to avoid generation of extra skyrmions under the large current density that complicates the subsequent analysis, after $t = 1.5$ ns, $j_e$ is reduced to be 0.03 MA cm$^{-2}$. This current density, could further drive the motion of the existing $Q = 0$ topologically trivial bubble, which would transform into an elongated shape. Following the expansion of the trivial bubble, the size of the $Q = +1$ antiskyrmion reduces accordingly. When the antiskyrmion shrinks to the lattice size, the antiparallel magnetization configuration abruptly switches to be parallel, namely, the antiskyrmion is annihilated at $t = 1.78$ ns. It is noteworthy mentioning that spin waves are excited and emitted from the position where the antiskyrmion is annihilated that results in energy dissipation in the FM layer. The annihilation process is due to the fact that antiskyrmion is an unstable elementary excitation in the given FM system that annihilates as a function of time due to Gilbert damping[55]. Once the antiskyrmion is annihilated, the topological charge of the remained elongated bubble domain (at $t = 2$ ns) can be calculated as $Q = -1$ skyrmion. This is a direct result of topological conservation. The transition of spin topology of magnetization textures from being topologically trivial bubbles $Q = 0$ to topologically nontrivial skyrmion $Q = -1$ in the FM layer is thus mediated by the creation and annihilation of an antiskyrmion. Subsequently, the generated skyrmion continues to move into wide area of the FM layer, and eventually shrinks to an energy-favorable circular shape ($t = 6$ ns). Note that in our micromagnetic simulation, the influence of defects is not considered. However, it should be noted that it has also been shown that magnetic defects can give rise to a skyrmion-antiskyrmion generation, which subsequently due to their opposite gyrotropic Magnus forces separate laterally[46]. In this instance DMI can also favor the stability of one (*e.g.*, skyrmions) over the other (*e.g.*, antiskyrmions).

**Experimental generation of skyrmion via nonmagnetic point contact.**

We have numerically discussed the feasibility of generating skyrmions underneath a nonmagnetic contact in the presence of inhomogeneous spin-orbit torques. Below we will realize it experimentally in a device made of a Ta (5 nm)/$Co_{20}Fe_{60}B_{20}$ (1.1 nm)/$TaO_x$ (3 nm) trilayer and a nonmagnetic point contact of Ti (10 nm)/Au (100 nm) as is shown in the schematic in Fig. 1(a). The multilayer was grown by ultrahigh vacuum magnetron sputtering and patterned via standard photo-lithography and ion milling. The nonmagnetic Ti/Au constriction is of length 20 μm and of width 4 μm. A polar magneto-optical Kerr effect (MOKE) microscopy was used for imaging the magnetic domain patterns.

Fig. 3 shows the creation and the subsequent translational motion of magnetic skyrmions. In the presence of a perpendicular magnetic field of $B_\perp = +0.5$ mT, only one isolated skyrmion bubble on the left side of the Ti/Au bridge can be seen. The magnetization on right side of trilayer remains in a positive saturation state, as shown in Fig. 3 (a). On the other hand, after passing a positive voltage pulse of amplitude 15 V (from left to right) and of duration 1 ms through the device, the generation of skyrmion bubbles is evident on the right side of Ti/Au constriction, as shown in Figs. 3 (b) – (d). The corresponding current density in the Ti/Au constriction is computed as 8.8 MA cm$^{-2}$ by measuring the resistance of device ($R =$ 3.86 kΩ), a current density that is comparable with the one used in micromagnetic simulation. It is noted that voltage/current is converted into electron current direction. The location of the skyrmion formation coincides with the contact region between the Ti/Au electrode and magnetic multilayer at which the divergent electron current and hence divergent SOTs are the largest. It is thus conceivable that the divergence of SOTs is critical for triggering magnetization instabilities that helps to form magnetic skyrmions[38,56]. After applying the pulse, the magnetization configuration on the right side of the device, however, remains the same as before. While the generation process of skyrmion bubbles can be understood based on the above micromagnetic simulation, the details cannot be experimentally identified here due to limitations in both the temporal and spatial resolution of the MOKE microscope. These skyrmion bubbles, once generated, bear a translational motion as a result of both inhomogeneous SOTs and gyrotropic Magnus force. Furthermore, upon applying consecutive voltage pulses, many additional skyrmion bubbles were generated and move into the wider area that indicates the reproducibility of the generation. This leads to the accumulation of skyrmions on the lower left part of the device, consistent with the skyrmion Hall effect.

It is known that the Magnus force can result in a skyrmion Hall effect and skyrmion accumulation in the presence of homogeneous SOTs[48,57]. The present system however, also involves contribution from the inhomogeneous current induced SOTs[11,16]. In order to elucidate whether the origin of skyrmion accumulation is from the Magnus force or from the spatially inhomogeneous SOTs, below we adopted a modified Thiele equation to analyze the dynamics of the magnetic skyrmion[11,29,31,58,59]:

$$\mathcal{G} \times \mathbf{v} - \alpha \mathcal{D} \cdot \mathbf{v} - 4\pi \mathcal{B} \cdot \mathbf{j}_e = 0 \qquad (2)$$

where the first term describes the Magnus force which results in the transverse motion of skyrmions with respect to the injected electron current $\mathbf{j}_e$, $\mathcal{G} = (0,0,-4\pi Q)$ is the gyromagnetic coupling vector and $\mathbf{v} = (v_x, v_y)$ is the drift velocity of the skyrmion. Note that the sign of the gyrotropic Magnus force is determined by both the sign of topological charge ($Q$) and the direction of motion of skyrmion ($\mathbf{v}$). The second term describes the dissipative force with the dissipative tensor $\mathcal{D}$ describing the effect of the drag force on the moving skyrmion $\mathcal{D} = 4\pi \begin{pmatrix} \mathcal{D}_{xx} & \mathcal{D}_{xy} \\ \mathcal{D}_{yx} & \mathcal{D}_{yy} \end{pmatrix}$. For a rigid Néel skyrmion (no distortion during motion), relations between each component are $\mathcal{D}_{xx} = \mathcal{D}_{yy} = \mathcal{D}$ and $\mathcal{D}_{xy} = \mathcal{D}_{yx} = 0$, and the value of $\mathcal{D} = \pi^2 d / 8\gamma_{dw}$ can be estimated based on the skyrmion profile[11], where $d$ is the diameter of skyrmion and $\gamma_{dw}$ is the domain wall width. The third term is the driving force provided by current-induced SOT with $\mathbf{j}_e = (j_x, j_y)$ and $\mathcal{B} = \frac{\tau_{ad}}{j_e} \begin{pmatrix} -\mathcal{J}_{xy} & \mathcal{J}_{xx} \\ -\mathcal{J}_{yy} & \mathcal{J}_{yx} \end{pmatrix}$ that quantifies the efficiency of the SOTs over the skyrmion bubbles. The components are given as $\mathcal{J}_{\mu\nu} =$ $\frac{1}{4\pi}\int\left(\frac{\partial m}{\partial \mu}\times m\right)_\nu dxdy$, where $\mu$ and $\nu$ run over $x$ and $y$. For the rigid hedgehog-like skyrmion, the relations that hold are $\mathcal{I}_{xx} = \mathcal{I}_{yy} = 0$ and $\mathcal{I}_{xy} = -\mathcal{I}_{yx} = \mathcal{I}$.

In our constricted device geometry, the current distribution $(j_x, j_y)$ can be computed by using Laplace's equation[40,44]. After solving the Eq. (2), we obtain the projection of the skyrmion velocities along the $x$- and $y$-directions as follows:

$$v_x = \left|\frac{\tau_{ad}}{j_e}\right|\mathcal{I}\frac{\alpha\mathcal{D}j_x+Qj_y}{Q^2+\alpha^2\mathcal{D}^2} \text{ and } v_y = \left|\frac{\tau_{ad}}{j_e}\right|\mathcal{I}\frac{\alpha\mathcal{D}j_y-Qj_x}{Q^2+\alpha^2\mathcal{D}^2}. \tag{3}$$

By defining a skyrmion Hall angle $tan\Phi_{sk} = v_y/v_x$, it is found that

$$tan\Phi_{sk} = \frac{\alpha\mathcal{D}j_y-Qj_x}{\alpha\mathcal{D}j_x+Qj_y} \tag{4}$$

Note that Eq.(4) can be further simplified as $\Phi_{sk} = -Q/\alpha\mathcal{D}$ when the driving currents (spin-orbit torques) are homogeneous as has been previously used for quantifying the skyrmion Hall effect in various multilayers[11,16].

From Eq. (4), it is clear that both the inhomogeneous current distribution $(j_x, j_y)$, the value of $\alpha\mathcal{D}$ and the spin topology of the skyrmion ($Q = \pm 1$) could influence the magnitude of skyrmion Hall angle. In the following, we will examine the contribution from each component individually. For the given material specific parameters, the domain wall widths can be calculated as $\gamma_{dw} = \pi\sqrt{\mathcal{A}/K_{eff}} \approx 63$ nm that leads to the value of $\mathcal{D} = \pi^2 d/8\gamma_{dw}$ to be determined as $\mathcal{D} \approx 20$. For the Ta/CoFeB/TaO$_x$ trilayer, the magnetic damping parameters is $\alpha \approx 0.02$. The value of $\alpha\mathcal{D}$ is estimated to be $\approx 0.4 < 1$. For the given triangular geometry, the current distribution $j_y/j_x$ around the contact region is about 1. Thus, the sign of skyrmion Hall angle is predominantly controlled by the sign of topological charge $Q$.

As indicated by Eq. (4), the sign of the gyrotropic Magnus force ($\mathbf{G} \times \mathbf{v}$) is governed by both the sign of topological charge ($Q = \pm 1$) and the direction of skyrmion motion ($\mathbf{v}$). The sign of topological charge can be reversed upon changing the polarity of the perpendicular magnetic fields since $Q = 1/4\pi\int \mathbf{m}\cdot\left(\partial\mathbf{m}/\partial x \times \partial\mathbf{m}/\partial y\right)dxdy$ is an odd function of magnetization vector $\mathbf{m}$. The direction of skyrmion motion, after moving away from the constricted region, is controlled solely by the (electron) current along the horizontal direction ($\pm x$). This enables the phase diagram of skyrmion Hall effect as a function of spin topology ($\pm Q$) and driving currents ($\pm j_e$) to be established, as shown in Fig. 4. By reversing the current/field directions, we can clearly see that the location of skyrmion accumulation is changed accordingly in our symmetrically designed device. In addition, the position of skyrmion generation, is consistent with the position where the divergence of SOTs is maximized. When the polarity of (electron) current pulse is fixed along the $-x$ direction (*i.e.*, $-j_e$), the position of skyrmion accumulation reverses from the top to the bottom of the left side of device after changing the perpendicular magnetic field from $B_\perp = -0.5$ mT to $B_\perp = +0.5$ mT. This is again consistent with the fact that the topological Magnus force changes its sign upon reversal of magnetic field (from $Q = +1$ to $Q = -1$). By reversing the polarity of (electron) current pulse to the $+x$ direction (*i.e.*, $+j_e$), the location of skyrmion accumulation is being reversed. These observations are consistent with the theoretical prediction and suggest that gyrotropic Magnus forces dominate the skyrmion dynamics.

**Conclusion**

In summary, we have numerically and experimentally studied the creation of magnetic skyrmions in a Ta/CoFeB/TaO$_x$ trilayer together with a nonmagnetic Ti/Au point contact. The topological transition of spin textures from being a topologically trivial bubble $Q = 0$ to a topologically nontrivial skyrmion $Q = -1$ is clearly resolved in micromagnetic simulations. These simulations also show the important role of skyrmion-antiskyrmion pairs in forming isolated skyrmion. Experimentally, in devices with a nonmagnetic point contact, we show skyrmions can be dynamically generated near the contact region. The skyrmion formation is directly related to the spatially divergent inhomogeneous current-induced SOTs. Since a nonmagnetic conducting Ti/Au electrode allows sufficient electric current to flow through, we further observed the spin-topology driven skyrmion dynamics, *i.e.*, the skyrmion Hall effect. The observation of the skyrmion Hall effect can be well described by a modified Thiele equation in the presence of spatially divergent SOTs. Our results could provide useful information for manipulating magnetic skyrmions in a controllable manner and for designing functional skyrmion devices that is based on the unique spin topology of magnetization textures.

**Methods**

The Ta(50 Å)/Co$_{20}$Fe$_{60}$B$_{20}$(CoFeB)(11 Å)/TaO$_x$(30 Å) trilayer was grown onto a semi-insulating Si substrate with 300-nm thick thermally formed SiO$_2$ layer, by using a dc magnetron sputtering technique. TaO$_x$ layer was prepared by oxidizing the top Ta layer via oxygen plasma of 10 W for 60 s and subsequently annealed in vacuum for 30 minutes. Devices were patterned by using standard photolithography and subsequent Ar ion milling. The Ti/Au point contacts were made following a lift-off process. The intrinsic magnetic parameters employed in our simulation are: exchange stiffness $A = 10 \times 10^{-12}$ J m$^{-1}$, magnetic anisotropy $K = 2.89 \times 10^5$ J m$^{-3}$ (effective anisotropy $K_{\text{eff}} = 2.34 \times 10^4$ J m$^{-3}$), DMI constant $D = 0.5 \times 10^{-3}$ J m$^{-2}$, $M_S = 6.5 \times 10^5$ A m$^{-1}$, $\gamma_0 = 2.211 \times 10^5$ m A$^{-1}$ s$^{-1}$, $\alpha = 0.1$, and $\theta_{sh} = -0.2$. The electrical resistivities of the electrode (Au), FM layer (CoFeB), and HM layer (Ta) are set as [53]: $\rho_{\text{Au}} = 2.3 \times 10^{-8}$ Ω m, $\rho_{\text{CoFeB}} = 1.7 \times 10^{-6}$ Ω m, $\rho_{Ta} = 1.9 \times 10^{-6}$ Ω m. The model is discretized into tetragonal cells of 5 nm × 5 nm × 1.1 nm and free tetrahedral elements with a maximum element size of 10 nm in the finite-difference micromagnetic simulation and the finite-element current distribution calculation, respectively. Polar magneto-optical Kerr effect images were acquired by using a commercial MOKE microscope from evico magnetics. For all images presented, a background image is subtracted. The background image is taken while applying a sufficiently large field to magnetically saturate the sample.

**ACKNOWLEDGEMENTS**

Work carried out at Tsinghua University was supported by the Basic Science Center Project of NSFC (Grant No. 51788104), National Key R&D Program of China (Grant Nos. 2017YFA0206200 and 2016YFA0302300), the NSFC Grant Nos. 11774194, 51831005, 1181101082, and the Beijing Advanced Innovation Center for Future Chip (ICFC). X.Z. acknowledges the support by the Presidential Postdoctoral Fellowship of the Chinese University of Hong Kong, Shenzhen (CUHKSZ). Y.Z. acknowledges the support by the President's Fund of CUHKSZ, the National Natural Science Foundation of China (Grant No. 11574137), and Shenzhen Fundamental Research Fund (Grant Nos. JCYJ20160331164412545 and JCYJ20170410171958839). X.L. acknowledges the support by the Grants-in-Aid for Scientific Research from JSPS KAKENHI (Grant Nos. 17K19074, 26600041 and 22360122). Work carried out at the Argonne National Laboratory (lithography, MOKE imaging, and Ti/Au deposition) was supported by the U.S. Department of Energy, Office of Science, Basic Energy Science, Materials Science and Engineering Division.

**Author contributions**

W.J. conceived and designed the experiments. G.Y. and K.W. fabricated the thin film. W.J., A.H. and S.G.E.te V. performed lithographic processing and MOKE imaging experiment. X.Z., J.X., X.L. and Y.Z. performed micromagnetic simulation. Z.W., L.Z., K.W. and W.J.

conducted data analysis. Z.W., X.Z. and W.J. wrote the manuscript. All authors commented on the manuscript.

**Additional information.**

Preprints and permission information is available online at www.nature.com/reprints. Correspondence and requests for materials should be addressed to W.J.

**Competing financial interests**

Authors declare no competing financial interests.

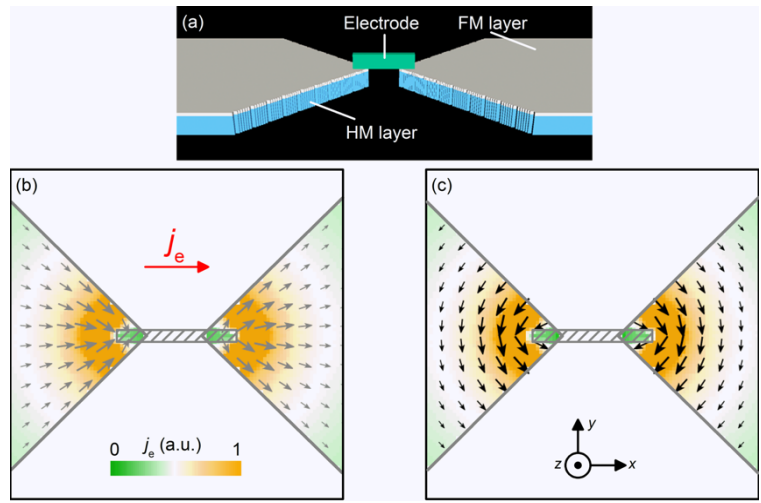

FIG. 1. (a) Schematic illustration of the geometrically constricted HM/FM heterostructure device which is connected through a nonmagnetic electrode. (b) The normalized electron current density distribution in the HM layer. (c) The corresponding spin current density distribution. The arrow indicates the direction, and the color as well as the arrow size denote the density.

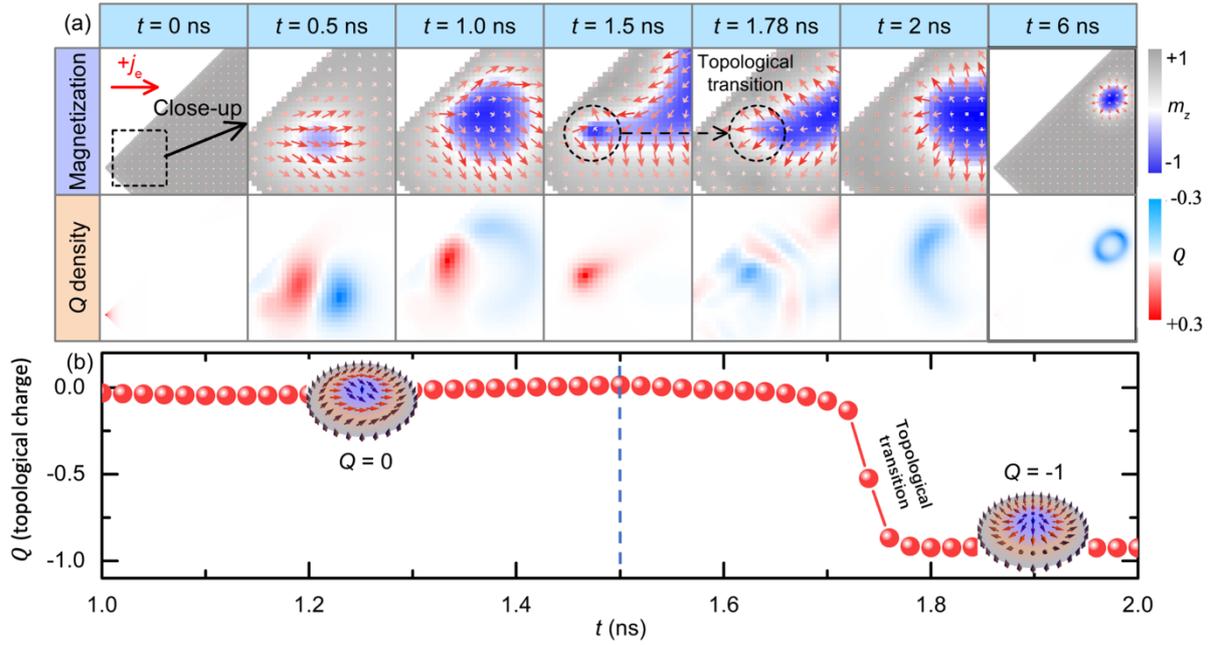

FIG. 2. Numerical demonstration of skyrmion generation enabled by inhomogeneous current induced SOTs around the nonmagnetic contact region. (a) Evolution of magnetization profile (upper panel) and the accompanied topological charge density profile (lower panel) as a function of time ($t$). Note that only part of spin textures is shown for time $t = 1.5$ ns, $t = 1.78$ ns and $t = 2$ ns, some spin textures with negative topological charge are out of the view window. (b) The calculated time dependent topological charge (i.e. skyrmion number $Q$) based on the varying magnetization configurations. The dashed line at $t = 1.5$ ns indicates the switching of current from 6 MA cm$^{-2}$ to 0.03 MA cm$^{-2}$. The transition of spin topology from $Q = 0$ to $Q = -1$ is evident around t = 1.7 ns.

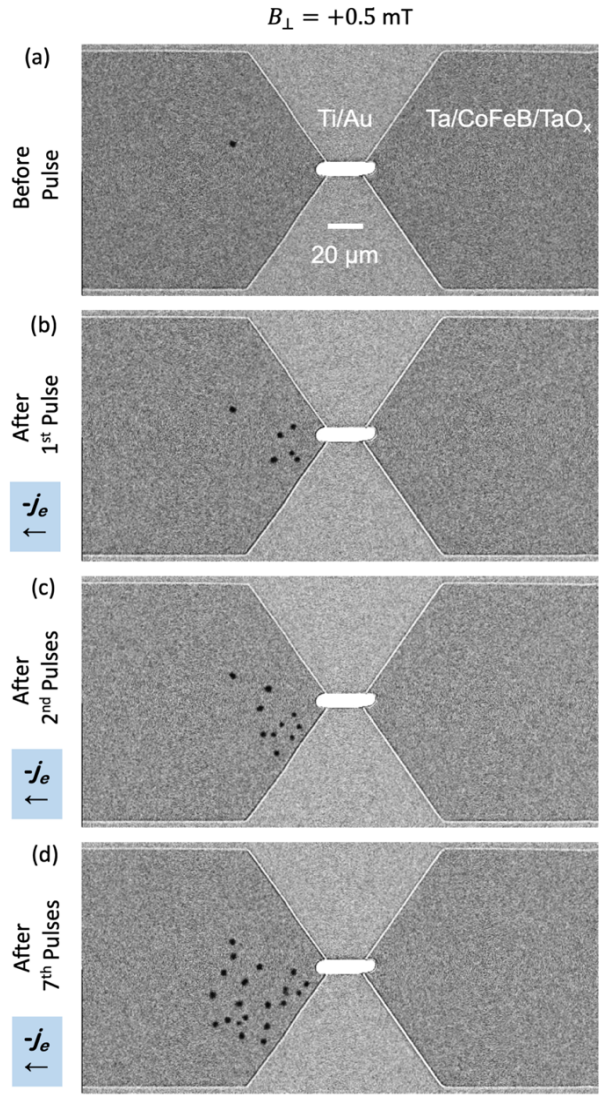

FIG. 3. Sequential MOKE images driven by a train of voltage pulses showing continuous creation and the subsequent accumulation of skyrmions at lower side of the devices, namely, the occurrence of skyrmion Hall effect. The experiment was performed under a perpendicular magnetic field $B_\perp = +0.5$ mT. The applied voltage pulses are of +15 V in amplitude and 1 ms in duration.

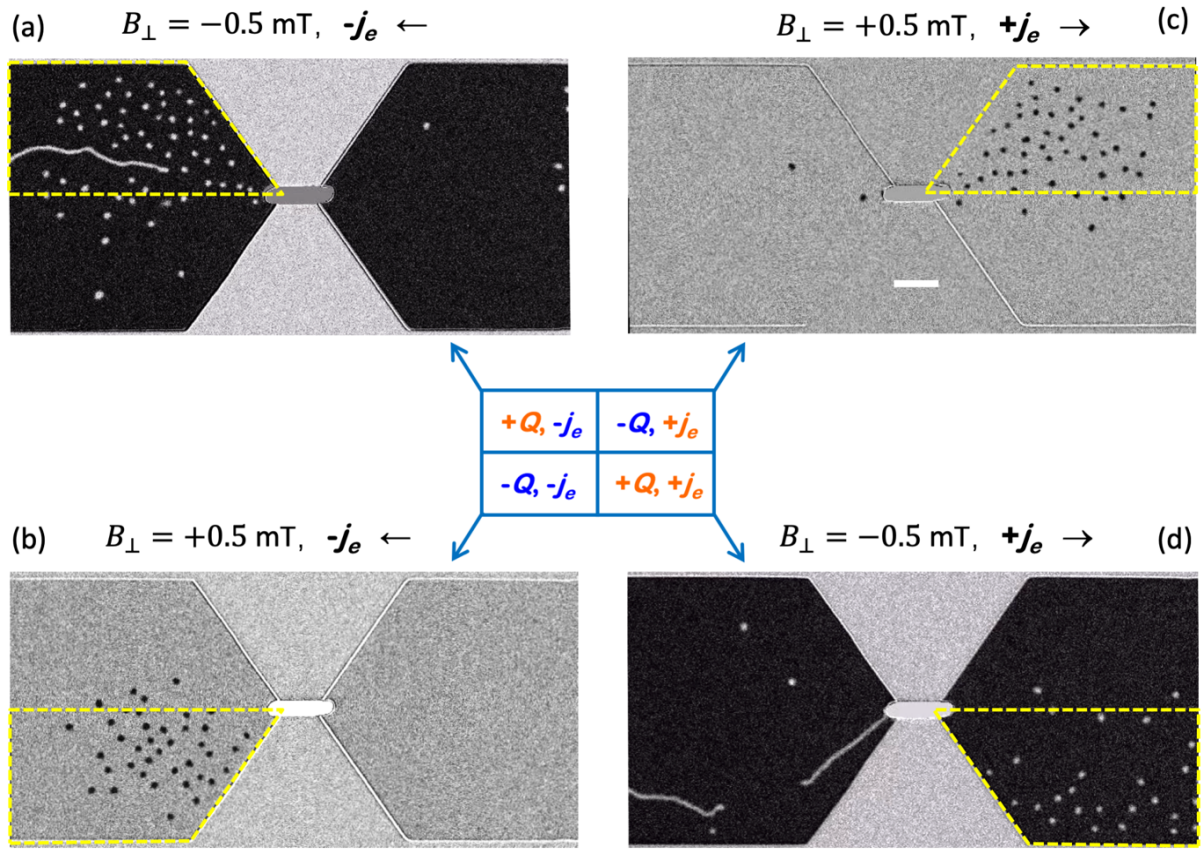

FIG. 4. Phase diagram of the skyrmion Hall effect for opposite sign of topological charge ($Q = \pm 1$) and opposite direction of voltage/current polarity ($\pm j_e$). (a) $-j_e$ flows toward the $-x$ direction with $Q = +1$ ($B_\perp = -0.5$ mT). (b) $-j_e$ flows toward the $-x$ direction with $Q = -1$ ($B_\perp = +0.5$ mT). (c) $+j_e$ flows toward the $+x$ direction with $Q = -1$ ($B_\perp = +0.5$ mT). (d) $j_e$ flows toward the $+x$ direction with $Q = +1$ ($B_\perp = -0.5$ mT). These images were taken after applying 20 pulses of +15 V in amplitude and 1 ms in duration. Scale bar corresponds to 20 μm.